\documentclass{article}
\usepackage{spconf,amsmath,graphicx}

\usepackage[table,xcdraw]{xcolor}

\usepackage{enumitem}
\setlist{nosep, leftmargin=14pt}

\usepackage{mwe} %

\title{Few-Shot Airway-Tree Modeling using Data-Driven Sparse Priors}
\name{Ali Keshavarzi, Elsa Angelini
}
\address{LTCI, Telecom Paris, Institut Polytechnique de Paris\\
 Palaiseau, France}
\begin{document}
\maketitle
\begin{abstract}
The lack of large annotated datasets in medical imaging is an intrinsic burden for supervised Deep Learning (DL) segmentation models. Few-shot learning approaches are cost-effective solutions to transfer pre-trained models using only limited annotated data. However, such methods can be prone to overfitting due to limited data diversity especially when segmenting complex, diverse, and sparse tubular structures like airways. Furthermore, crafting informative image representations has played a crucial role in medical imaging, enabling discriminative enhancement of anatomical details. In this paper, we initially train a data-driven sparsification module to enhance airways efficiently in lung CT scans. We then incorporate these sparse representations in a standard supervised segmentation pipeline as a pretraining step to enhance the performance of the DL models. Results presented on the ATM public challenge cohort show the effectiveness of using sparse priors in pre-training, leading to segmentation Dice score increase by 1\% to 10\% in full-scale and few-shot learning scenarios, respectively.
\end{abstract}
\begin{keywords}
Airway segmentation, Few-shot learning, Sparse Representation
\end{keywords}

\vspace{-1em}
\section{Introduction}
\label{sec:introduction}
\vspace{-0.5em}

Automated airway modeling from lung CT scans greatly aids in the diagnosis and prognosis of pulmonary diseases such as bronchiectasis, Chronic Obstructive Pulmonary Disease (COPD), and airway wall thickening. An airway tree is a branching network of tubular, tapered-shaped structures with various sizes and orientations from the trachea, to main, segmental and distal bronchi.  

The development of automated airway-tree segmentation depends on the availability of large annotated datasets, which are difficult to access and costly to generate. 
The emergence of few-shot learning methods has high potential impact for this application.

As a machine learning paradigm, few-shot learning enables pre-trained models to specialize on a new task with a small number of training samples.
However, developing robust few-shot pipelines to avoid overfitting remains challenging.
We believe that a key factor for efficient few-shot learning in the context of airway segmentation is the robust construction of image representations. 

Before the emergence of DL methods, most airway and vessel extraction algorithms relied on pre-filtering to enhance structures of interest~\cite{LESAGE2009819}. Hessian-matrix-based methods such as Frangi (vesselness) filters~\cite{hessian_frangi}, were among the most popular approaches for characterizing the local geometry of image information and enhance tubular structures prior to the segmentation task. 
While the vesselness filter has been used for multiple segmentation tasks and imaging modalities ~\cite{LESAGE2009819, vesselness_liver, vesselness_examples_2}, it has two major limitations: high computational complexity and poor enhancement performance at bifurcations. 
As an alternative, data-driven approaches such as texton dictionary learning have also been quite popular, able to learn image-specific visual characteristics from the data itself and optimized for sparse encoding. Sparse coding~\cite{sparse_coding_3rd} has been popular for tasks like image restoration~\cite{image_restoration_elad}, image denoising~\cite{sparse_repr_denoising_elad}, medical image fusion~\cite{sparse_fusion_liu}, and image classification~\cite{ZHANG201744}.

Recently, DL models have dominated the medical image segmentation field and achieved significant results that outperform conventional approaches, especially with encoder-decoder convolutional architectures~\cite{unet_origi}. DL-based airway segmentation has been developed over the past few years~\cite{garcia_3DUNet_post, wingsnet}. 
However, common issues with these models include unknown optimal network depth, huge networks with numerous hyperparameters, the need for large annotated datasets, and long convergence time. For example, recent popular architectures like AttentionUNet~\cite{attentionunet} and WingsNet~\cite{wingsnet} have demonstrated notable performance on airway segmentation tasks, yet their large size and training time remain an issue for replication and further fine-tuning on new cohorts (e.g. new scanners) or new diseases, especially with limited number of annotated cases. 

\begin{figure*}[hbt!]
    \centering
    \label{fig:cdl_csc_pipeline}
    \includegraphics[width=0.95\textwidth]{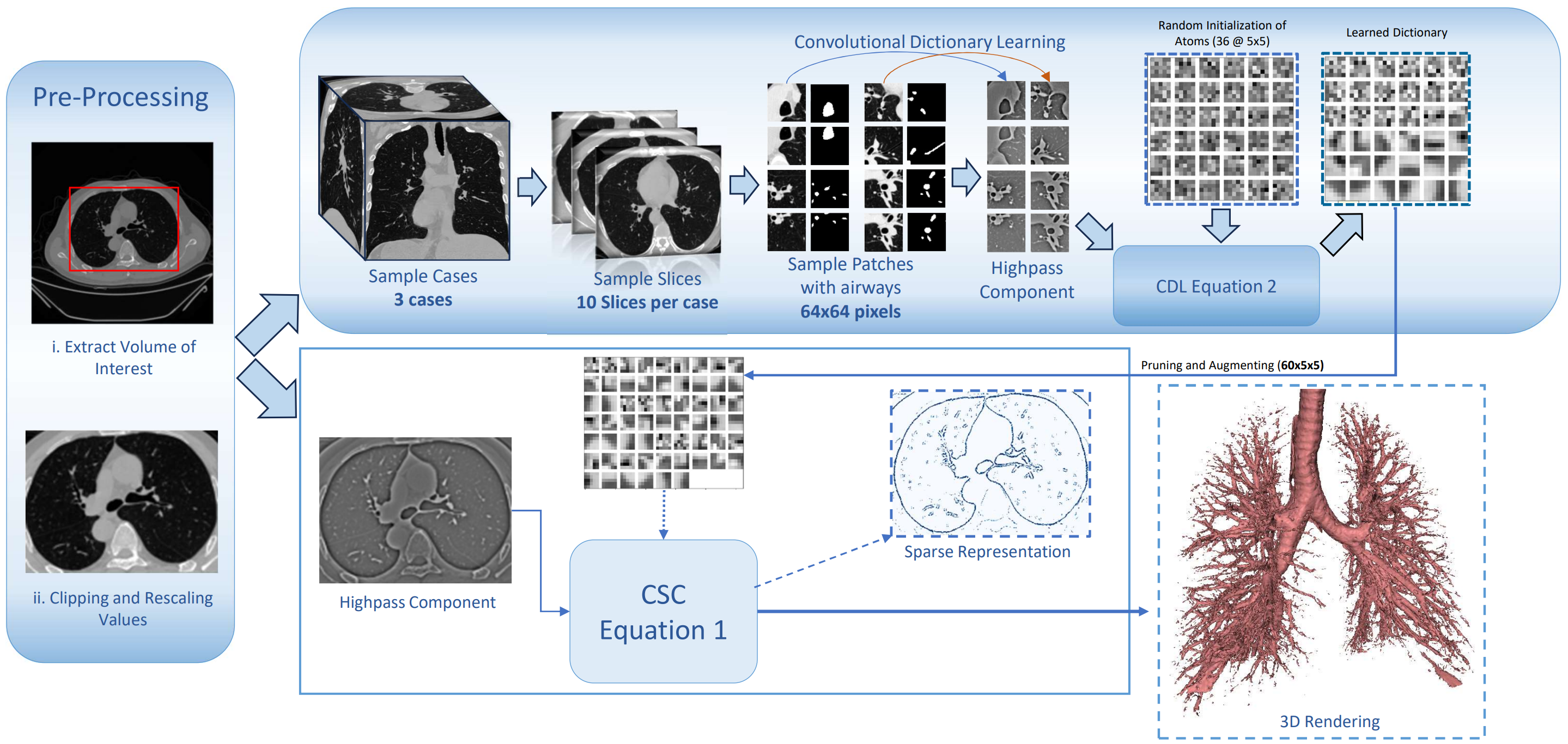}
    \caption{CDL and CSC pipeline for creating sparse representations on lung CT scans.}
    
\end{figure*}

Our main contributions to collectively address the challenges associated with airway-tree segmentation in low-resource settings are summarized as follows:
\begin{itemize}
  \item We introduce an efficient, fast, and practical few-shot learning airway segmentation model. %
  \item We explore the efficacy of incorporating sparse representations in the segmentation of airways in lung CT scans as a powerful image enhancement tool to capture and focus on visual characteristics of airways. %
  \item Our experiments show a significant drop in the training time for our few-shot learning setup using sparse priors while preserving good performance in airway segmentation tasks. 
\end{itemize}

\section{Methodology}
\label{sec:methodology}

\subsection{Convolutional Sparse Coding}
\label{subsec:method_csc}
Convolutional Sparse Coding (CSC) is a recent framework enabling to learn a dictionary of convolutional filters to sparsely encode a specific type of data. In our case, CSC is trained to encode an image as the sum of convolutions over coefficient maps with the same size as the source image and correspond to a learned set of dictionary filters. CSC is more complex and sophisticated than classic sparse coding, allowing the filters to recognize local spatial structures in the input data~\cite{effcsc_wohlberg}. For an input image s, CSC encoding is expressed as: 

\begin{align}
    \label{eq:csc}
    \underset{\{x_{k}\}}{\arg \min}  \, \frac{1}{2}  \left \| \underset{k}{\sum} d_{k} \ast x_k - s \right \|_{2} ^{2} + \lambda \underset{k}{\sum} \left \| x_{k} \right \|_{1},
\end{align}
where \{$d_{k}$\} is the set of $K$ learned dictionary filters, $\ast$ denotes convolution, and \{$x_{k}$\} is the set of coefficient maps. If the dictionary of filters is unknown, the CSC encoding can be formulated as a Convolutional Dictionary Learning (CDL) problem as follows:

\begin{equation}
    \begin{split}
    \label{eq:cdl}
    \underset{\{d_k\},\{x_{k}\}}{\arg \min} \; \frac{1}{2} \; \left \| \underset{k}{\sum} d_k \ast x_{k} \; - \; s \right \|_2^2 + \lambda \underset{k}{\sum} \left \| x_{k} \right \| _{1},
    \end{split}
\end{equation}
\vspace{3em}
such that $ d_k \in C \; \forall k$; where $C$ is the feasible set of filters with unit norm and constrained support and $\lambda$ is the sparsity regularization term in both CSC and CDL.

In our work, CSC representations of lung CT scans were learned and generated using the SPORCO~\cite{wohlberg-2017-sporco} open-source library. It offers not only a set of adaptable and efficient algorithms for CSC encoding and dictionary learning but also for visualizing and analyzing the results.
Using the Alternating Direction Method of Multipliers (ADMM) Wohlberg~\cite{effcsc_wohlberg} algorithm, the dictionary learning stage in SPORCO alternates between sparse coding and dictionary learning update stages. The hyperparameters\footnote{The hyperparameters for the CDL and CSC were selected through an exhaustive search.} for the dictionary learning (Eq. \ref{eq:cdl}) are initialized as follows: initial dictionary $d_k$ size = 36 atoms, of 5x5 pixels with zero-mean and $\lambda=0.2$. 

Figure~\ref{fig:cdl_csc_pipeline} depicts the steps involved in applying CDL and CSC algorithms to lung CT axial slices. The CT scans are cropped to the volume of interest (VoI) around the lungs and intensity values are clipped in the range $[-1000, 600]$ HU and then rescaled to $[0, 1]$ range. A set of 50 patches of size 64x64 is extracted on random axial slices from 3 CT scans with the constraint of including airways (based on ground truth annotations in our case).

The CDL algorithm is then applied to the highpass component of the patches obtained using the Tikhonov filter. The learned filters $d_k$ are then pruned so that we only keep the atoms whose corresponding coefficient map $x_{k,j}$ is non-zero. Augmenting the dictionary involves up-down/left-right flips and rotation of the atoms by 90 degrees. Finally, using the learned, pruned, and augmented dictionary filters, the CSC algorithm is applied to the clipped, rescaled highpass component of the VoI to obtain a sparse representation of any lung CT volume. 

\begin{figure}[!h]
\begin{minipage}[b]{0.24\linewidth}
  \centering
  \centerline{\includegraphics[width=\textwidth]{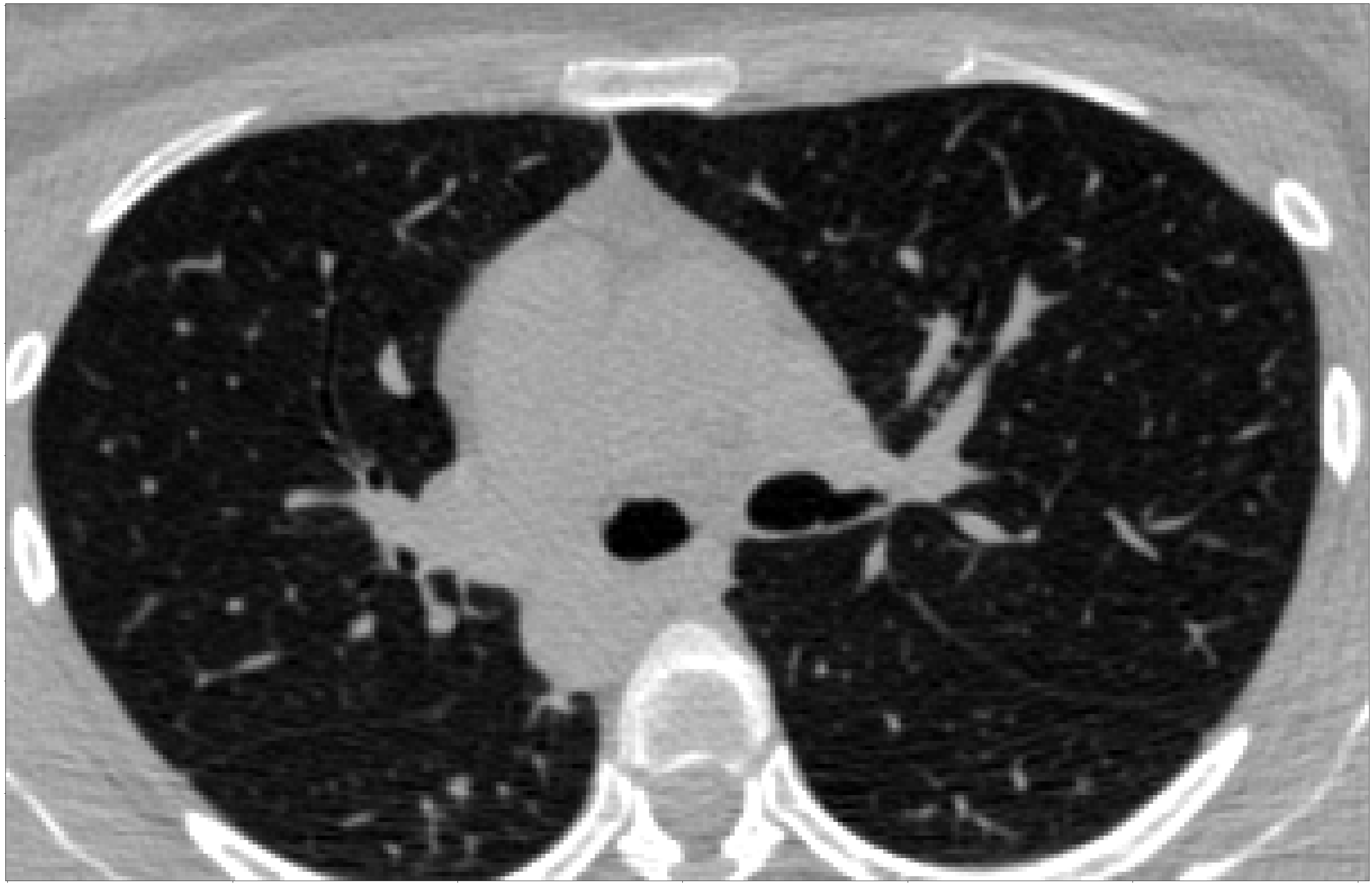}}

  \centering
  \centerline{\includegraphics[width=\textwidth]{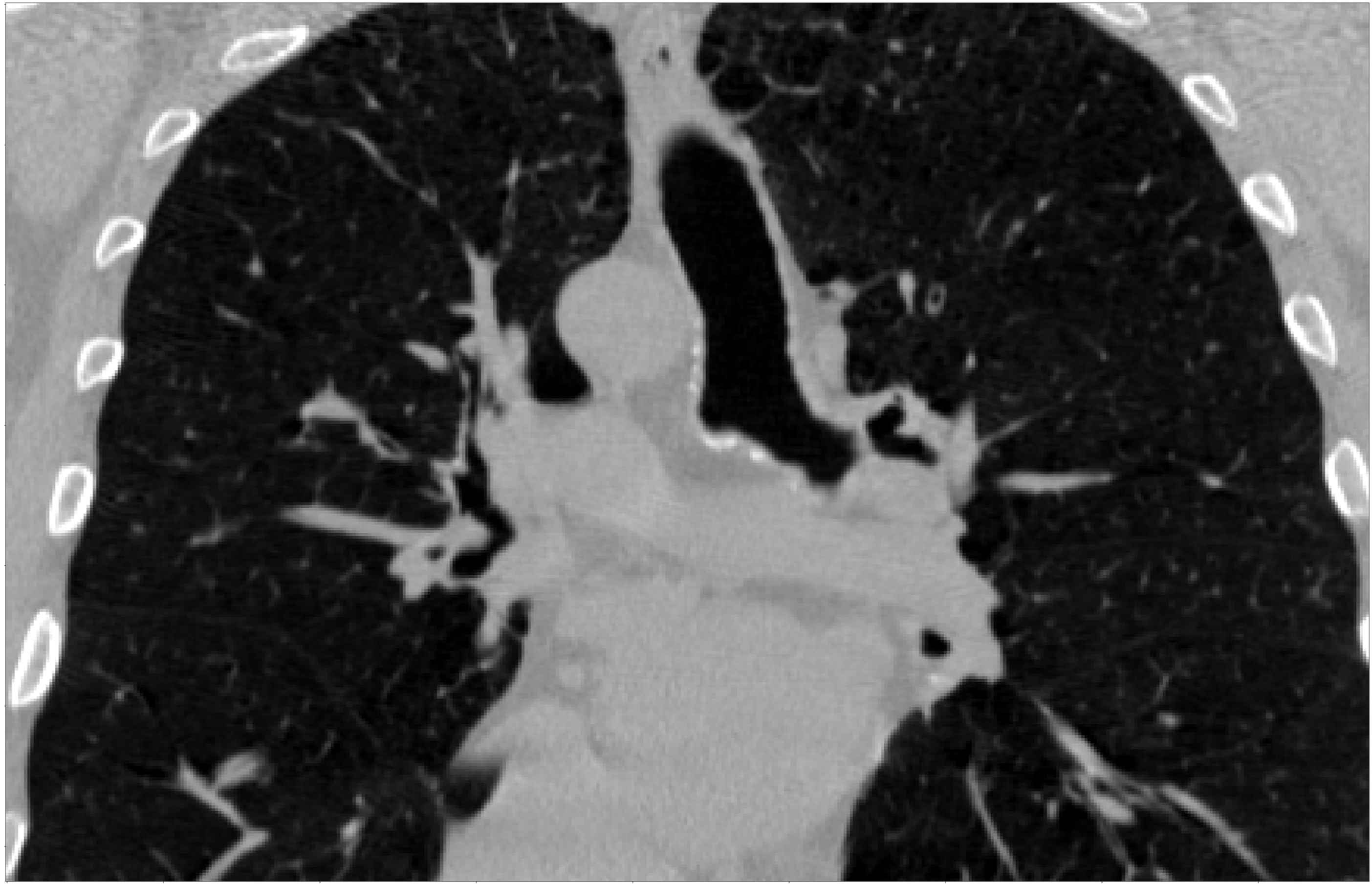}}

  \centerline{Original CT}\medskip
\end{minipage}
\hfill
\begin{minipage}[b]{0.24\linewidth}
  \centering
  \centerline{\includegraphics[width=\textwidth]{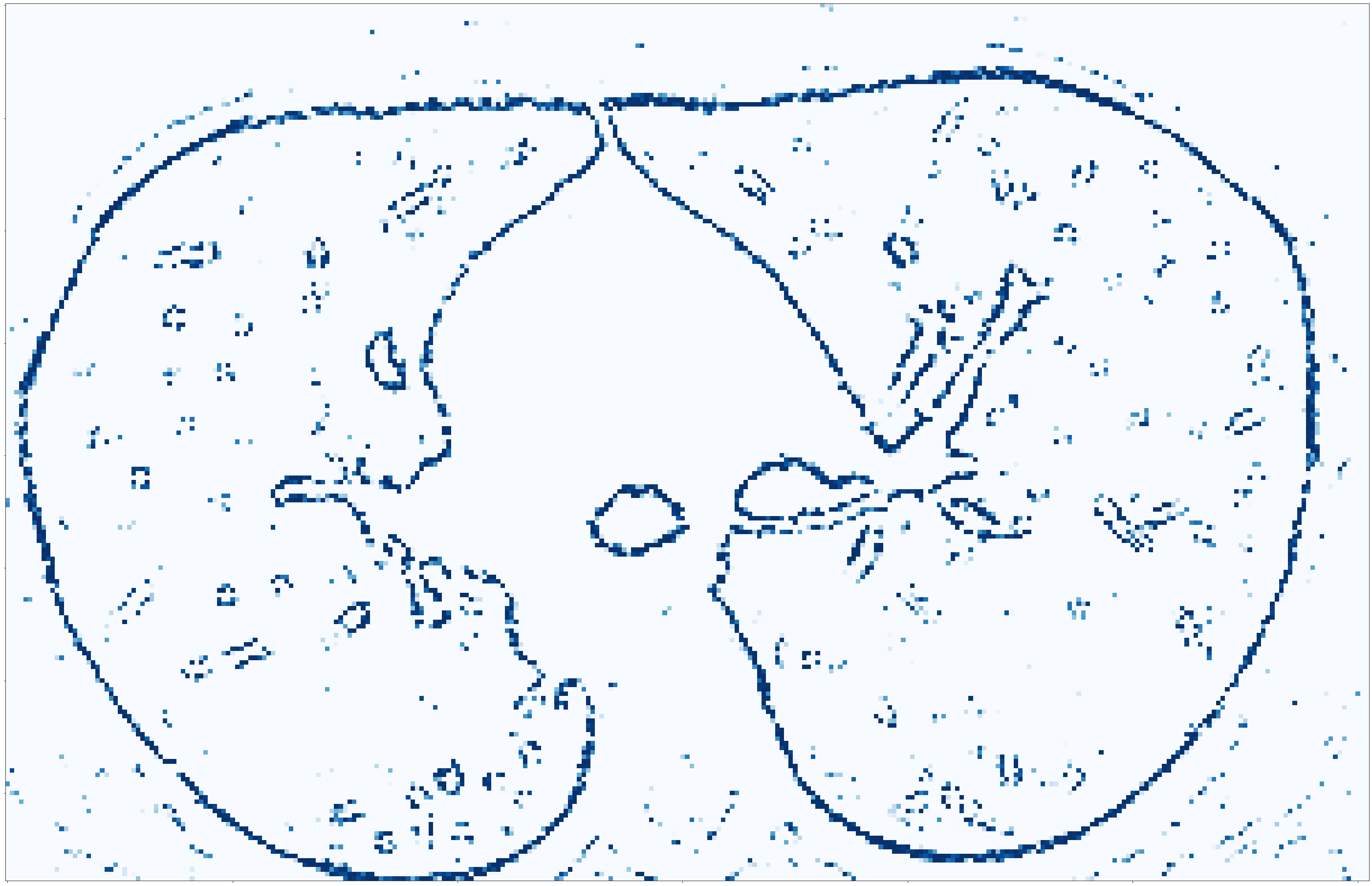}}

  \centering
  \centerline{\includegraphics[width=\textwidth]{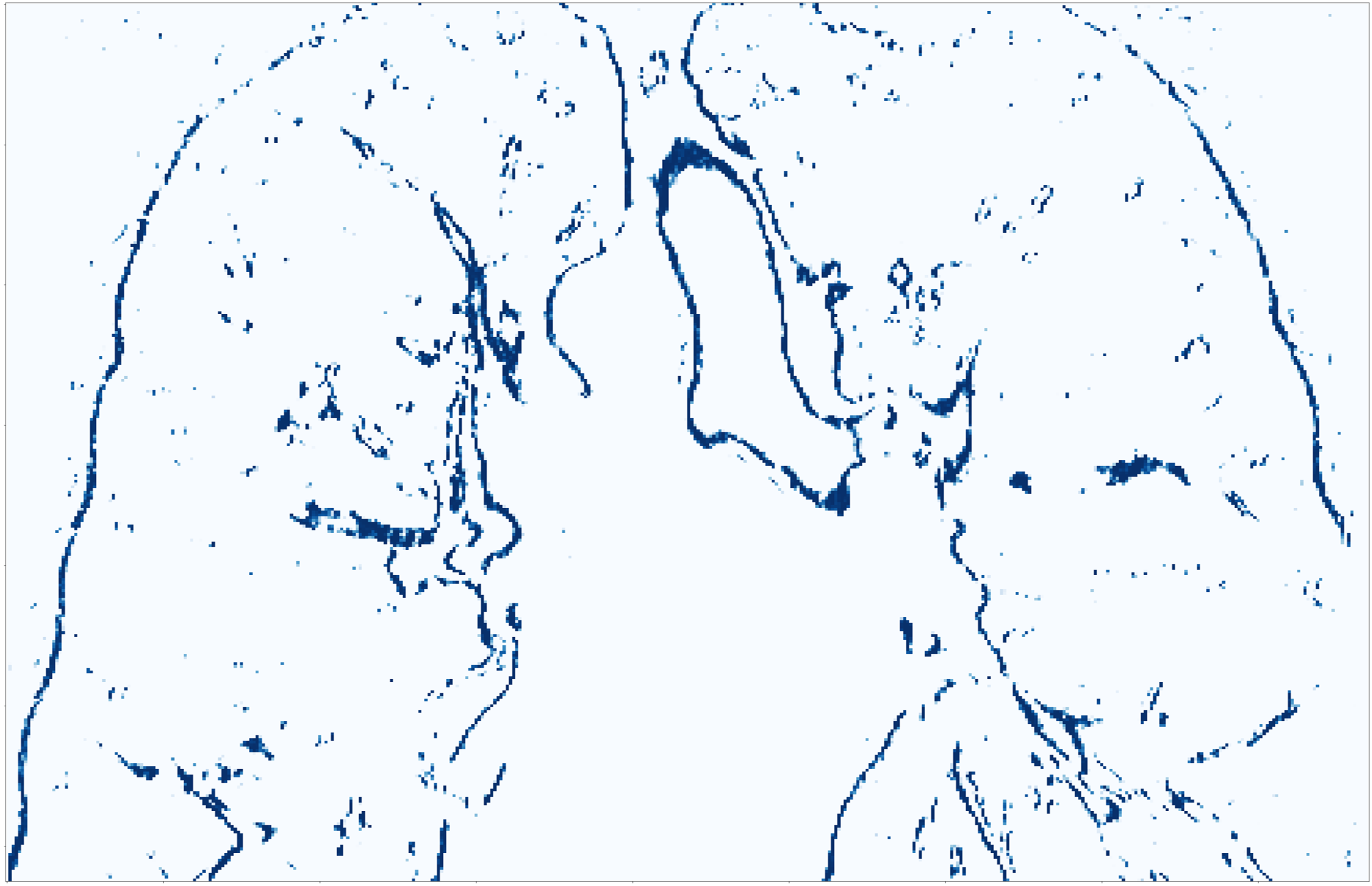}}

  \centerline{Sparse}\medskip
\end{minipage}
\hfill
\begin{minipage}[b]{0.24\linewidth}
  \centering
  \centerline{\includegraphics[width=\textwidth]{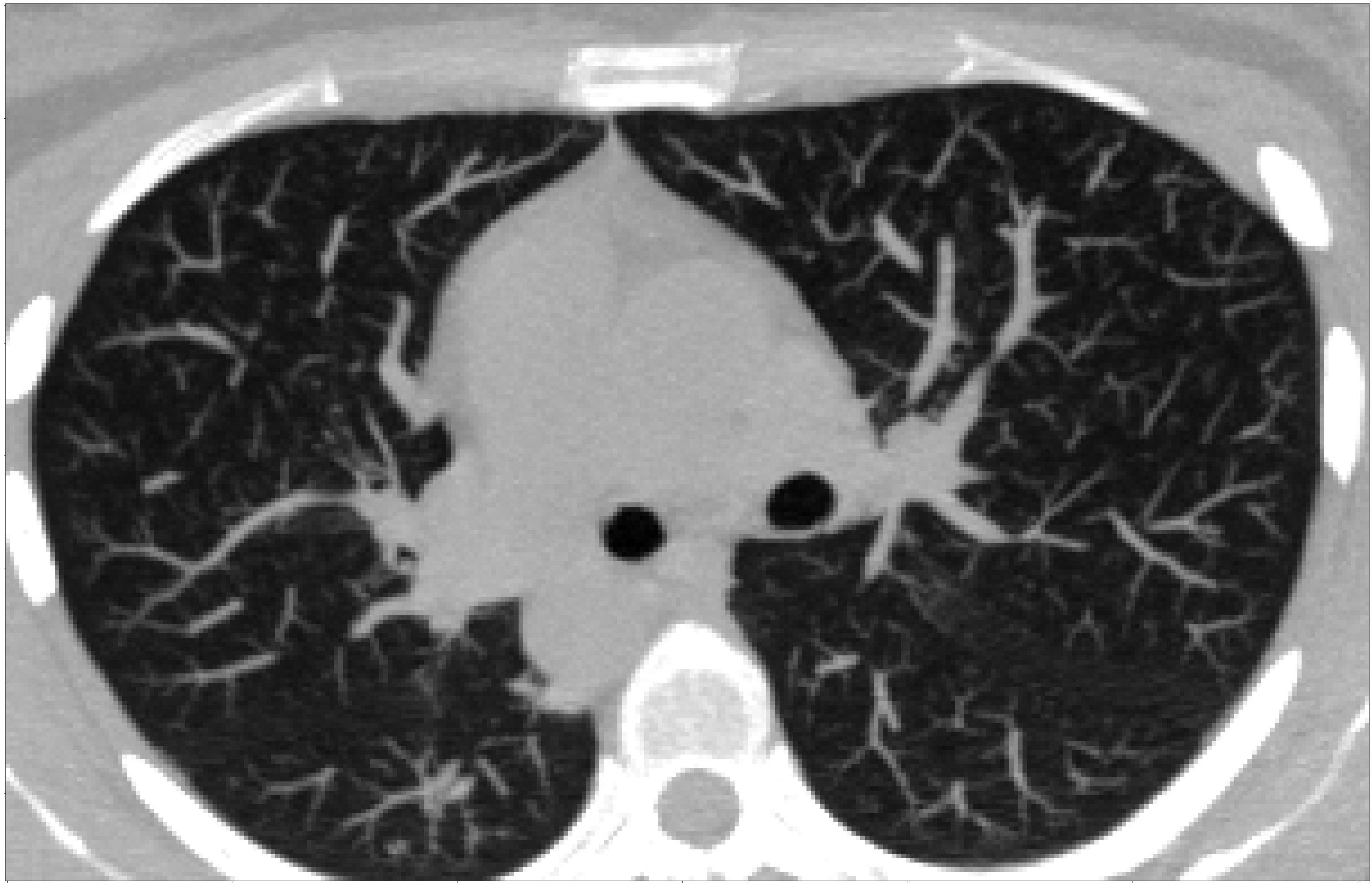}}

  \centering
  \centerline{\includegraphics[width=\textwidth]{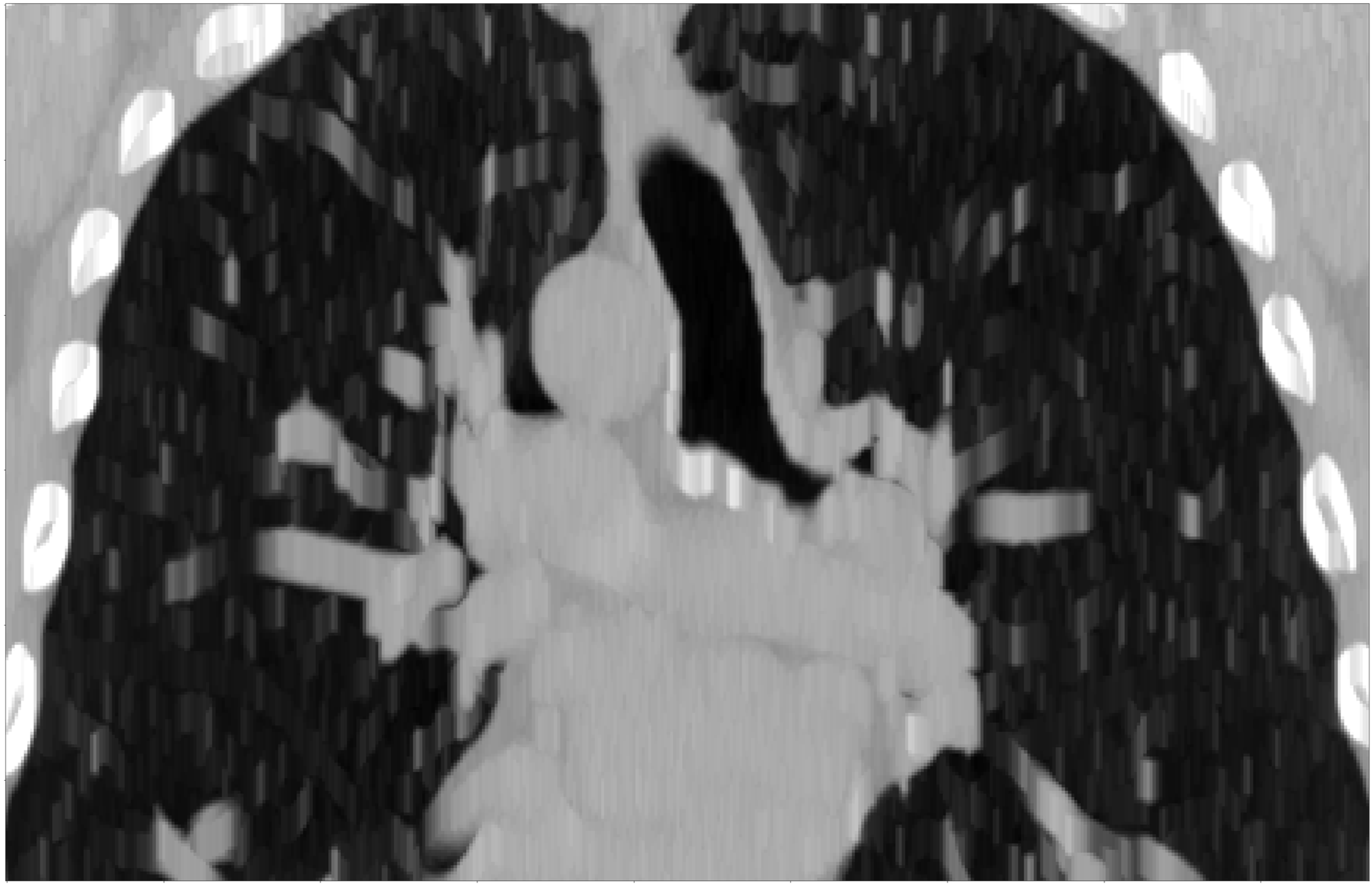}}

  \centerline{MIP}\medskip
\end{minipage}
\hfill
\begin{minipage}[b]{0.24\linewidth}
  \centering
  \centerline{\includegraphics[width=\textwidth]{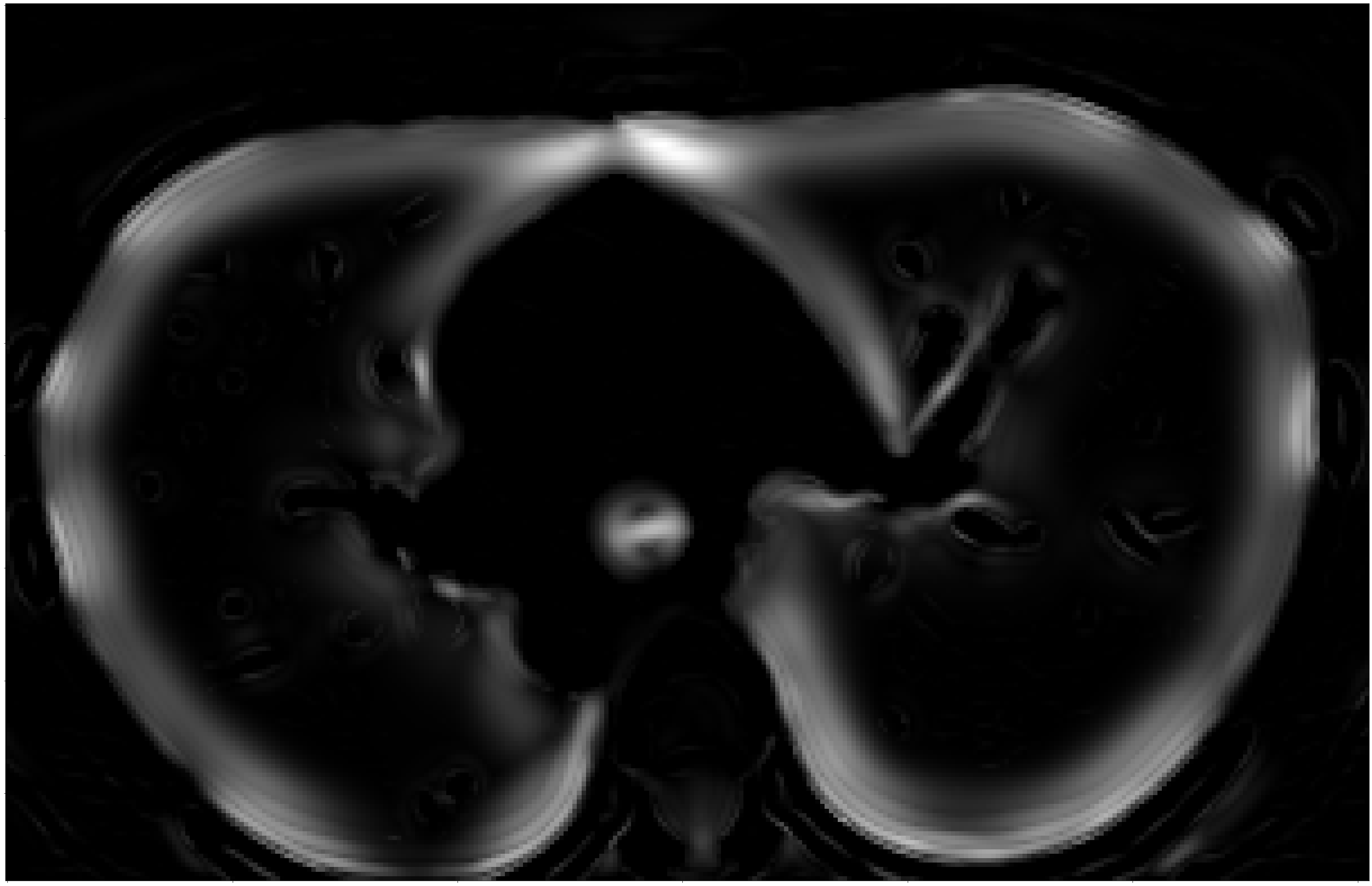}}

  \centering
  \centerline{\includegraphics[width=\textwidth]{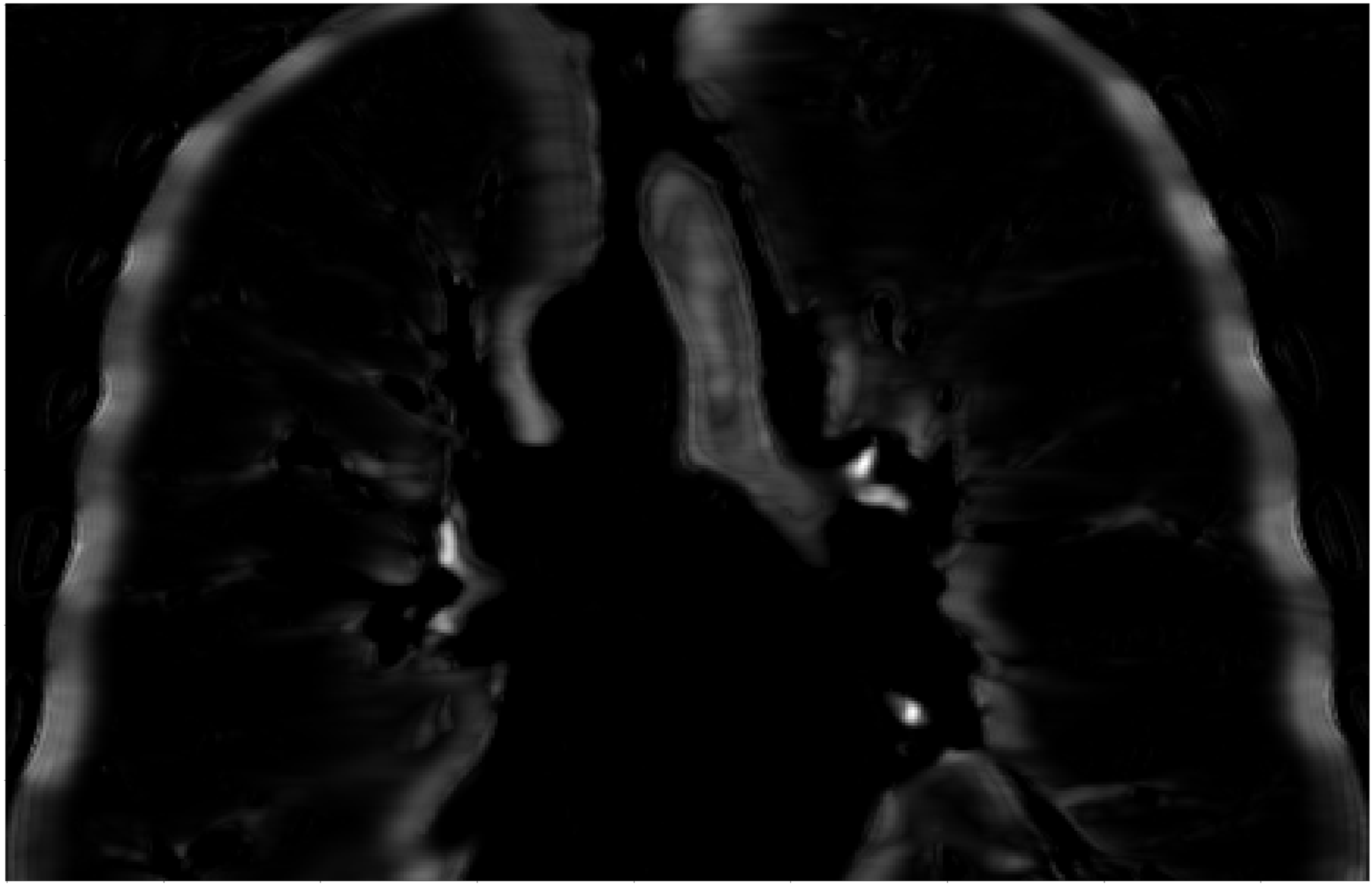}}

  \centerline{Vesselness}\medskip
\end{minipage}
\caption{Comparison between CSC encoding, MIP, and Vesselness filter on one axial and one coronal slice. 
}
\label{fig:mip_sparse}
\end{figure}

The efficiency of CSC in enhancing important structures such as airway edges, branching points, and high contrast regions is shown in Figure \ref{fig:mip_sparse} with a comparison to Maximum Intensity projection (MIP) and Vesselness filter~\cite{hessian_frangi}, which are two popular tubular structures enhancing methods.

\subsection{Segmentation Network Architectures}
 In our experiments, we used AttentionUNet \cite{attentionunet} as the segmentation model to perform pixel-level supervised binary segmentation on large 320x320x320 patches. The architecture employs 3D convolutions, along with Attention Gates (AGs) to improve the encoding-decoding process and force the model to focus on important areas by calculating attention weights for each feature map channel. 

The network is trained using the Dice coefficient as the loss function. The Dice loss function is a commonly used similarity-based loss function in image segmentation tasks, which calculates the ratio of overlap between the predicted and ground truth segmentation masks, and penalizes the network for low overlap. The Dice loss can be defined as follows:

\begin{align}
    \label{eq:dice_loss}
    D = 1 - \frac{2\times \sum_{i=1}^{N} p_i g_i}{\sum_{i=1}^{N} p_i + \sum_{i=1}^{N} g_i},
\end{align}
where $p_i$ is the prediction probability of the voxel and $g_i$ is the corresponding ground truth label.

\subsection{Pretraining and Few-shot learning}
The network was pre-trained or fully trained on combinations of original and CSC encoded cases, as detailed in the results.  
When using pre-training, the trained weights are used as the initial configuration for fine tuning. For few shot learning with pre-training, the same small set of CT scans is used. 

\vspace{-1em}
\section{Experiments and Results}
\label{sec:exp_results}

\subsection{Training and Implementation Details}

\label{subsec:training_imp_details}

The implementation of the 3D attention U-Net architecture is composed of five layers, with feature dimensions of size [16, 32, 64, 128, 256], and a stride value of 2.
Throughout all of our experiments, we have maintained a batch size of one, which means that each iteration of our model processes a single sample. In our implementation, we use Adam as the optimizer and we have set the initial learning rate to 1e-3 and the weight decay to 1e-5. The type of scheduling procedure of our pipeline is ReduceLROnPlateau with a patience of 15 epochs with a factor of 0.1. All of our experiments were executed using a single NVIDIA A100-SXM4-80GB GPU.
In the few-shot pretraining experiment, CSC sparse representations of five cases were randomly chosen for network pretraining.
The same test set served as a consistent benchmark for all experiments, allowing us to fairly compare results.

\subsection{Dataset and Preprocessing}
\label{subsec:data_prepro}
The ATM22~\cite{zhang2023multi} dataset includes N=500 CT scans (300 for training, 150 for testing, and 50 for validation). While annotations are provided for the 300~\footnote{Case ATM\_164\_0000 has a misaligned label, therefore excluded from our experiments.} training scans, the publisher has kept annotations for the validation and the whole test sets private. 
In our experiments, the ATM22 public training set was consistently divided into 254 training cases and 45 test cases, with the test set remaining constant across all of the experiments.
Each CT scan contains 84 to 1,125 axial slices with a spatial size of 512x512 pixels. Voxel sizes vary from 0.51mm to 0.91mm in the axial plane and 0.5 to 5.0mm in slice thickness. Average in-plane pixel size = 0.77mm and average slice thickness = 0.56mm. CT scans come from different scanners and segmentations are provided for the trachea, main bronchi, lobar bronchi, and distal segmental bronchi,
Although resampling reduces computational costs, can improve the signal-to-noise ratio, and facilitates the integration of multiple imaging modalities, it disturbs the morphological properties of the airway structures, especially on small branches. We therefore avoided resampling in our pipeline.

Due to the large range of intensity values in lung CT scans and the fact that we are not interested in the outside of the lung parenchyma and other super-bright organs and tissues, it is necessary to clip the values. 
In the ATM22 dataset, the voxel values inside the ground-truth airway segmentations range from -1024 to -800 HU (with few values outside this range). We chose to clip intensity values to $[-1000, +600]$ HU.
To aid in learning and convergence, the HU values are then rescaled between 0 and 1. The LungMask~\cite{Hofmanninger2020} module was used to extract the Volume of Interest (VoI) from lung regions.

\subsection{Segmentation Results}
In Table \ref{tab:results}, the initial three rows display ours results with full training, while the subsequent three rows present the results in our few-shot learning setting. The training time for each experiment is shown in the fourth column of Table \ref{tab:results}.

Our baseline fully trained on the training set of ATM22 has a Dice score of 95.5. 
The same network architecture, pre-trained on the sparse encoded scans and then fine-tuned on the original images improves the Dice score to 96.4. 
and reduces average fine-tuning time by a few hours.

\begin{table}[h!]
\centering
\begin{tabular}{cccc}
\hline
\textbf{Pretraining}              & \textbf{Fine-tuning}        & \textbf{Dice}           & \textbf{Time}                \\ \hline
\multicolumn{4}{c}{ \textbf{Full Training set}}   \\    \hline
- & ATM22           & 95.5 & 18.05 h. \\ \hline
- & $\text{ATM22}_{sp}$     & 93.0 & 14.9 h. \\ \hline
$\text{ATM22}_{sp}$    & ATM22     &\textbf{96.4} & \textbf{9.3 h.} \\ \hline
\multicolumn{4}{c}{ \textbf{Few-Shot - Mini Training set}}   \\    \hline
- & MiniATM22 & $80.3 \pm 7.1$  & 1.2 h.  \\ \hline
- & $\text{MiniATM22}_{sp}$ & $84.1 \pm 1.3$  & 40 min. \\ \hline
$\text{MiniATM22}_{sp}$ & MiniATM22 & $90.2 \pm 1.4$ & \textbf{22 min.} \\ \hline
\end{tabular}

\caption{Segmentation results on the test set using either the full training set or a mini set of 5 randomly selected CT scans, and with and without sparse encoding $_{sp}$ of the CT scans. Few-shot learning experiments were run with 10 different mini sets.
}
\label{tab:results}
\end{table}
When the model is only trained on the MiniATM22 and $\text{MiniATM22}_{sp}$ datasets, the Dice score favors the sparse encoded data but drops to 84.1. A noticeable improvement is achieved when we pre-train on $\text{MiniATM22}_{sp}$ and then fine-tune on miniATM22, with a Dice score at 90.2 and significant reduction in average fine-tuning time. Moreover, the reported Dice score standard deviation over different training samples for few-shot learning is reduced when using the sparse encoding. 

\begin{figure}[htb!]
\begin{minipage}[b]{0.31\linewidth}
  \centering
  \centerline{\includegraphics[width=\textwidth]{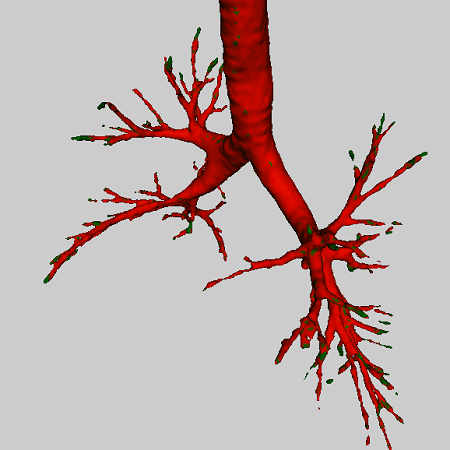}}
  \centerline{baseline}\medskip

  \centering
  \centerline{\includegraphics[width=\textwidth]{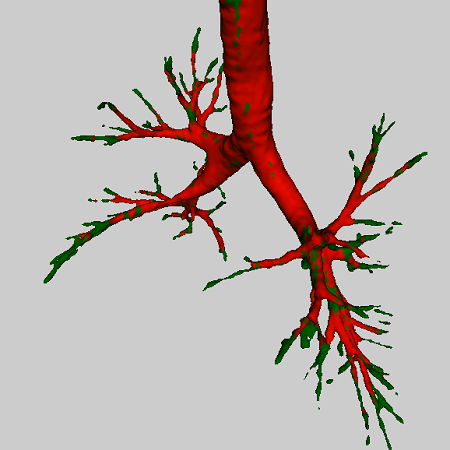}}
  \centerline{FS}\medskip

  \centering
  \centerline{\includegraphics[width=\textwidth]{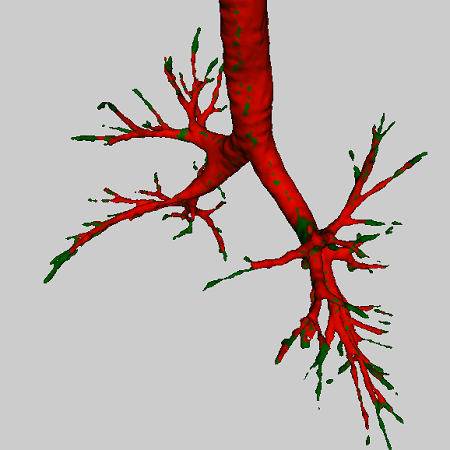}}
  \centerline{FS + Pre-train}\medskip

\end{minipage}
\hfill
\begin{minipage}[b]{0.31\linewidth}
  \centering
  \centerline{\includegraphics[width=\textwidth]{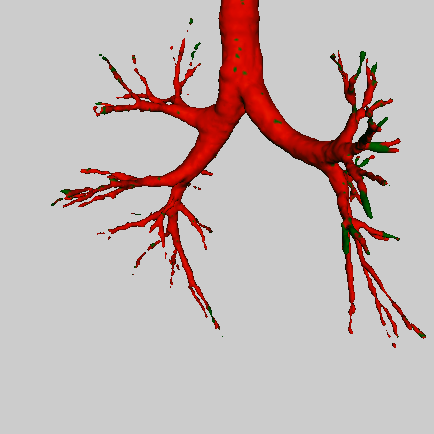}}
  \centerline{baseline}\medskip

  \centering
  \centerline{\includegraphics[width=\textwidth]{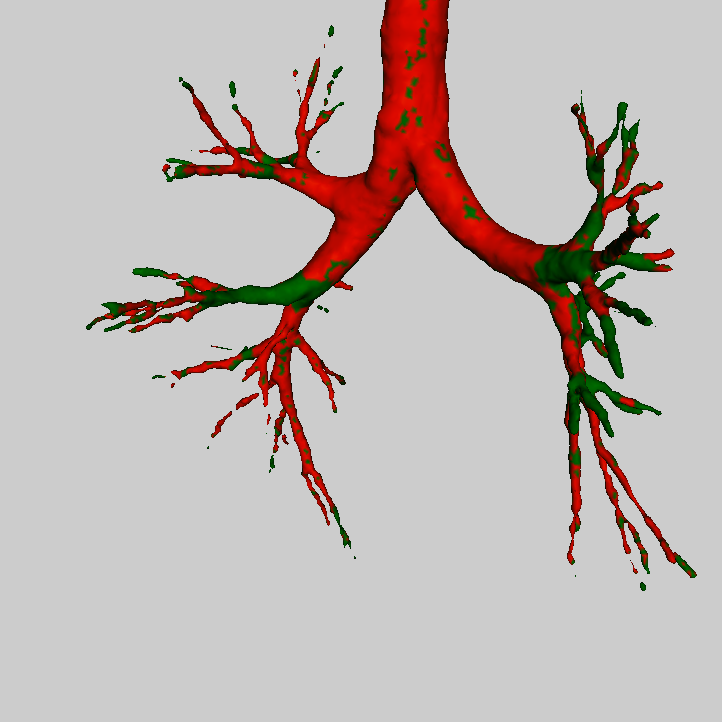}}
  \centerline{FS}\medskip

  \centering
  \centerline{\includegraphics[width=\textwidth]{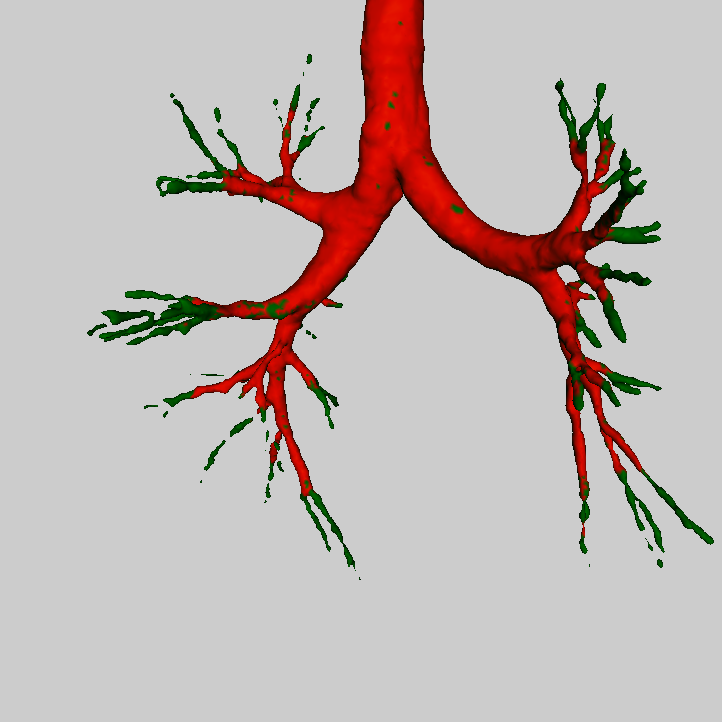}}
  \centerline{FS + Pre-train}\medskip
  
\end{minipage}
\hfill
\begin{minipage}[b]{0.31\linewidth}
  \centering
  \centerline{\includegraphics[width=\textwidth]{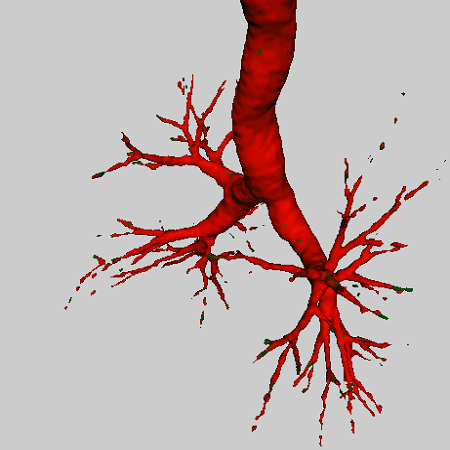}}
  \centerline{baseline}\medskip

  \centering
  \centerline{\includegraphics[width=\textwidth]{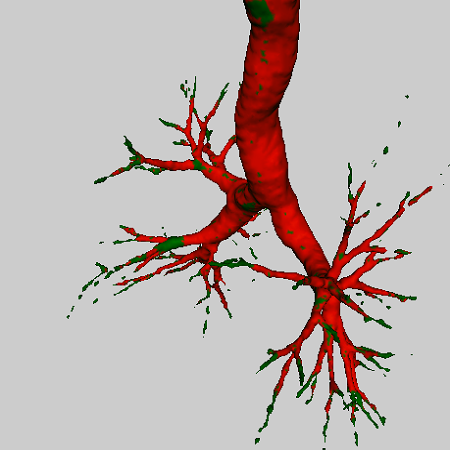}}
  \centerline{FS}\medskip

  \centering
  \centerline{\includegraphics[width=\textwidth]{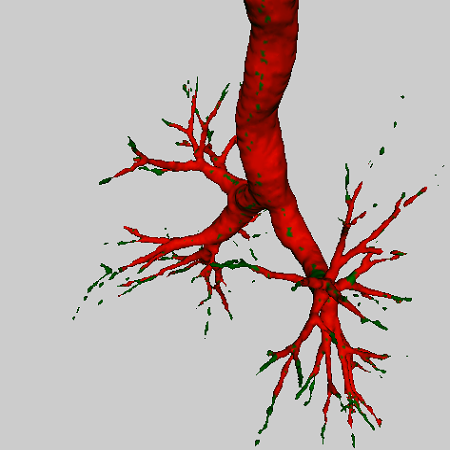}}
  \centerline{FS + Pre-train}\medskip

\end{minipage}

\caption{3D reconstruction of segmented airway trees on 3 cases: GT (green) and predictions (red); First row: results from full training on the whole training cohort ATM22 (baseline). Second and third rows: results from few-shot (FS) learning on original CTs in MiniATM22 (2nd row) and with pre-training on $\text{MiniATM22}_{sp}$ (3rd row). 
}
\label{fig:res-3D}
\end{figure}

We visually illustrated some results in Fig. \ref{fig:res-3D}. 
In the few-shot learning results, we can clearly appreciate higher performance on smaller branches and distal airways when pre-training with $\text{MiniATM22}_{sp}$. Notably, the second column shows a challenging segmentation case within our test set, as evidenced by the few-shot models' inability to perform well even on large branches. Nonetheless, our proposed method, which involves pretraining on sparse representations, outperforms the basic few-shot segmentation model significantly.

\section{Conclusion}
\label{sec:conclusion}
We proposed a novel paradigm for efficient supervised training of a UNet type of DL architectures to segment airways on lung CT scans. We showed that data-driven sparse encoding of lung CT scans enables few-shot learning with high Dice scores and training time within an hour. We also showed that pre-training on sparse encoded images leads to improved Dice in both full training and few-shot learning scenarios. Our convolutional sparse coding was specifically trained to encode visual features associated with airways. It is therefore very effective in capturing small airway details and discriminating airways from other structures.
Pre-training is becoming a major focus in the context of federated learning.
Future work will extend this framework for other cohorts and disease cases.

\section{acknowledgements}
\label{sec:ackno}
This research work is funded by the Doctoral School of IP-Paris and Hi!Paris and was performed using HPC resources from GENCI-IDRIS (Grant 2023-AD011013999).

\bibliographystyle{IEEEbib}
\bibliography{refs}

\begin{thebibliography}{10}

\bibitem{LESAGE2009819}
D.~Lesage, E.~Angelini, I.~Bloch, and G.~Funka-Lea,
\newblock ``A review of 3d vessel lumen segmentation techniques: Models, features and extraction schemes,''
\newblock {\em Medical Image Analysis}, vol. 13, no. 6, pp. 819--845, 2009.

\bibitem{hessian_frangi}
A.~F. Frangi, W.~J. Niessen, K.~L. Vincken, and M.~A. Viergever,
\newblock ``Multiscale vessel enhancement filtering,''
\newblock in {\em Medical Image Computing and Computer-Assisted Intervention --- MICCAI'98}, 1998, pp. 130--137.

\bibitem{vesselness_liver}
J.~Lamy, O.~Merveille, B.~Kerautret, N.~Passat, and A.~Vacavant,
\newblock ``Vesselness filters: A survey with benchmarks applied to liver imaging,''
\newblock in {\em 25th International Conference on Pattern Recognition (ICPR)}, 2021, pp. 3528--3535.

\bibitem{vesselness_examples_2}
A.~Longo, S.~Morscher, J.~Malekzadeh Najafababdi, D.~Jüstel, C.~Zakian, and V.~Ntziachristos,
\newblock ``Assessment of hessian-based frangi vesselness filter in optoacoustic imaging,''
\newblock {\em Photoacoustics}, vol. 20, pp. 100200, 2020.

\bibitem{sparse_coding_3rd}
Julien Mairal, Francis Bach, and Jean Ponce,
\newblock ``Sparse modeling for image and vision processing,''
\newblock {\em Foundations and Trends® in Computer Graphics and Vision}, vol. 8, no. 2-3, pp. 85--283, 2014.

\bibitem{image_restoration_elad}
Julien Mairal, Michael Elad, and Guillermo Sapiro,
\newblock ``Sparse representation for color image restoration,''
\newblock {\em IEEE Transactions on Image Processing}, vol. 17, no. 1, pp. 53--69, 2008.

\bibitem{sparse_repr_denoising_elad}
M.~Elad and M.~Aharon,
\newblock ``Image denoising via learned dictionaries and sparse representation,''
\newblock in {\em IEEE Computer Society Conference on Computer Vision and Pattern Recognition (CVPR)}, 2006, vol.~1, pp. 895--900.

\bibitem{sparse_fusion_liu}
Yu~Liu, Xun Chen, Aiping Liu, Rabab~K. Ward, and Z.~Jane Wang,
\newblock ``Recent advances in sparse representation based medical image fusion,''
\newblock {\em IEEE Instrumentation and Measurement Magazine}, vol. 24, no. 2, pp. 45--53, 2021.

\bibitem{ZHANG201744}
R.~Zhang, J.~Shen, F.~Wei, X.~Li, and A.~K. Sangaiah,
\newblock ``Medical image classification based on multi-scale non-negative sparse coding,''
\newblock {\em Artificial Intelligence in Medicine}, vol. 83, pp. 44--51, 2017.

\bibitem{unet_origi}
O.~Ronneberger, P.~Fischer, and T~Brox,
\newblock ``U-net: Convolutional networks for biomedical image segmentation,''
\newblock in {\em Medical Image Computing and Computer-Assisted Intervention -- MICCAI}, 2015, pp. 234--241.

\bibitem{garcia_3DUNet_post}
A.~Garcia-Uceda, R.~Selvan, Z.~Saghir, H.~A. W.~M. Tiddens, and M.~de~Bruijne,
\newblock ``Automatic airway segmentation from computed tomography using robust and efficient 3-d convolutional neural networks,''
\newblock {\em Scientific Reports}, vol. 11, no. 1, pp. 16001, Aug 2021.

\bibitem{wingsnet}
Hao Zheng, Yulei Qin, Yun Gu, Fangfang Xie, Jie Yang, Jiayuan Sun, and Guang-Zhong Yang,
\newblock ``Alleviating class-wise gradient imbalance for pulmonary airway segmentation,''
\newblock {\em IEEE Transactions on Medical Imaging}, vol. 40, no. 9, pp. 2452--2462, 2021.

\bibitem{attentionunet}
O.~Oktay, J.~Schlemper, L.~Le Folgoc, M.~Lee, M.~Heinrich, K.~Misawa, K.~Mori, S.~McDonagh, N.~Y. Hammerla, B.~Kainz, B.~Glocker, and D.~Rueckert,
\newblock ``Attention u-net: Learning where to look for the pancreas,''
\newblock in {\em Medical Imaging with Deep Learning}, 2018.

\bibitem{effcsc_wohlberg}
Brendt Wohlberg,
\newblock ``Efficient algorithms for convolutional sparse representations,''
\newblock {\em IEEE Transactions on Image Processing}, vol. 25, no. 1, pp. 301--315, 2016.

\bibitem{wohlberg-2017-sporco}
Brendt Wohlberg,
\newblock ``{SPORCO}: {A} {P}ython package for standard and convolutional sparse representations,''
\newblock in {\em 15th Python in Science Conference}, July 2017, pp. 1--8.

\bibitem{zhang2023multi}
Minghui Zhang, Y.~Wu, H.~Zhang, Y.~Qin, H.~Zheng, W.~Tang, C.~Arnold, C.~Pei, P.~Yu, Y.~Nan, et~al.,
\newblock ``Multi-site, multi-domain airway tree modeling,''
\newblock {\em Medical Image Analysis}, p. 102957, 2023.

\bibitem{Hofmanninger2020}
J.~Hofmanninger, F.~Prayer, J.~Pan, S.~R{\"o}hrich, H.~Prosch, and G.~Langs,
\newblock ``Automatic lung segmentation in routine imaging is primarily a data diversity problem, not a methodology problem,''
\newblock {\em Eur. Radiol. Exp.}, vol. 4, no. 1, pp. 50, Aug. 2020.

\end{thebibliography}

\end{document}